\documentclass[fleqn,usenatbib]{mnras}

\usepackage[T1]{fontenc}

\usepackage{enumitem}

\DeclareRobustCommand{\VAN}[3]{#2}
\let\VANthebibliography\thebibliography
\def\thebibliography{\DeclareRobustCommand{\VAN}[3]{##3}\VANthebibliography}

\usepackage{graphicx}	
\usepackage{amsmath}	
\usepackage{amssymb}	
\usepackage[dvipsnames]{xcolor}
\usepackage{physics}
\usepackage{natbib}
\usepackage{booktabs}
\usepackage{cleveref}
\usepackage{array}
\usepackage{multirow}
\usepackage{comment}
\usepackage{orcidlink}
\usepackage{bm}
\usepackage{enumitem}
\usepackage{calc}
\crefname{chapter}{Chapter}{Chapters}
\Crefname{chapter}{Chapter}{Chapters}

\crefname{equation}{equation}{equations}
\Crefname{equation}{Equation}{Equations}

\crefname{figure}{Fig.}{Fig.}
\Crefname{figure}{Figure}{Figures}

\crefname{table}{Table}{Tables}
\Crefname{table}{Table}{Tables}

\crefname{section}{Section}{Sections}
\Crefname{section}{Section}{Sections}

\usepackage{newtxtext,newtxmath}  

\newcommand{\diff}{\mathrm{d}}

\newcommand{\LCDM}{$\Lambda$CDM}

\newcommand{\oDE}{\omega_{\mathrm{DE}}}

\newcommand{\oK}{\omega_{K}}
\newcommand{\ob}{\omega_{\rm b}}
\newcommand{\oc}{\omega_{\rm c}}

\newcommand{\sotto}{\sigma_{8}}
\newcommand{\sottom}{\sigma_{8,\mathrm{m}}}
\newcommand{\sottoc}{\sigma_{8,\mathrm{c}}}
\newcommand{\sdodici}{\sigma_{12}}
\newcommand{\PL}{P_{\rm L}(k, z)}
\newcommand{\Pdd}{P_{\delta\delta}(k)}
\newcommand{\Pdt}{P_{\delta\theta}(k)}
\newcommand{\Ptt}{P_{\theta\theta}(k)}
\newcommand{\Pttanddt}{$\Ptt$ and $\Pdt$}
\newcommand{\Gadget}{\textsc{Gadget-4}}

\newcommand{\Mpch}{h^{-1} \, \mathrm{Mpc}}
\newcommand{\hMpc}{h \, \mathrm{Mpc}^{-1}}
\newcommand{\iMpc}{\, \mathrm{Mpc}^{-1}}
\newcommand{\Mpc}{\, \mathrm{Mpc}}

\newcommand{\kmax}{k_\mathrm{max}}
\newcommand{\kcrop}{k_\mathrm{crop}}
\newcommand{\shapep}{\Theta_{\rm s}}
\newcommand{\evolp}{\Theta_{\rm e}}

\newcommand{\sv}{\sigma_{v}}

\newcommand{\Pdw}{P_{\rm dw}(k, z)}
\newcommand{\Pnw}{P_{\rm nw}(k, z)}
\newcommand{\parsigma}{S}
\newcommand{\parmu}{C}

\defcitealias{Bel_2019}{B19}
\defcitealias{Jennings2012}{J12}
\DeclareRobustCommand{\Bel}{\citetalias{Bel_2019}}

\newcolumntype{C}{>{\centering\arraybackslash\hspace{4pt}}c<{\hspace{4pt}}}

\setlist[itemize]{leftmargin=20pt,
        itemindent=5pt,
        labelsep=5pt,
        labelwidth=25pt,
        labelindent=!}

\title[Improved recipes for velocity statistics]{Improved recipes for peculiar velocity power spectra using Evolution Mapping}

\author[M. Esposito et al.]{
Matteo Esposito$^{1}$\thanks{esposito@mpe.mpg.de}\orcidlink{0000-0001-6386-0997},
Ariel G. S\'anchez$^{1}$\orcidlink{0000-0003-1198-831X},
Julien Bel$^{2}$, and 
Andr\'es N. Ruiz$^{3,4}$
\\~\\
$^{1}$Max-Planck-Institut f\"ur Extraterrestrische Physik, Postfach 1312, Giessenbachstr., D-85748 Garching, Germany \\
$^{2}$Aix Marseille Univ, Université de Toulon, CNRS, CPT, Marseille, France\\
$^{3}$Instituto de Astronomía Teórica y Experimental (CONICET-UNC), Laprida 854, X5000BGR, Córdoba, Argentina\\
$^{4}$Observatorio Astronómico, Universidad Nacional de Córdoba, Laprida 854, X5000BGR, Córdoba, Argentina
}

\date{Accepted XXX. Received YYY; in original form ZZZ}

\pubyear{202X}

\begin{document}
\label{firstpage}
\pagerange{\pageref{firstpage}--\pageref{lastpage}}
\maketitle

\begin{abstract}
We present new fitting functions for the velocity divergence auto- and cross-power spectra, $P_{\theta\theta}(k)$ and $P_{\delta\theta}(k)$, calibrated on gravity-only $N$-body simulations. By applying the Evolution Mapping framework, we revise existing prescriptions to introduce a physically motivated parametrisation in terms of the clustering amplitude $\sigma_{12}$, the RMS density fluctuation smoothed at $12\,\text{Mpc}$. This approach improves robustness and extends the range of applicability beyond that of previous models.
Our fits are calibrated using a suite of multi-resolution simulations, with numerical convergence carefully quantified and sampling artefacts mitigated through a conservative patching strategy. This yields converged measurements up to $k\simeq0.56\,\mathrm{Mpc}^{-1}$ and percent-level accuracy for both $P_{\theta\theta}(k)$ and $P_{\delta\theta}(k)$ over a wide range of $\sigma_{12}$. Validation against independent simulations spanning a broad range of cosmological models confirms an accuracy of $1$–$2$ per cent on scales where the measurements are robust, systematically outperforming existing prescriptions.
We further assess the impact of deviations from the exact evolution mapping relation induced by differing growth histories. For most cosmologies of practical interest, we find that neglecting these effects introduces only subdominant errors. 
We show that expressing fitting functions in $h$-dependent units leads to spurious, unphysical dependencies on the Hubble parameter, even for models with identical linear clustering. This provides strong empirical support for parametrising non-linear evolution in terms of $\sigma_{12}$ rather than $\sigma_{8}$.
Our fitting functions provide a robust description of velocity power spectra, with direct applications to redshift-space distortion modelling in galaxy redshift surveys.
\end{abstract}


\section{Introduction}\label{sec:intro}
Current and upcoming galaxy surveys, including the Dark Energy Survey \citep[DES,][]{DES_constraints}, the Dark Energy Spectroscopic Instrument \citep[DESI,][]{DESI_DR1}, Euclid \citep{Euclid}, and the Vera Rubin Observatory Legacy Survey of Space and Time \citep[LSST,][]{LSST}, are mapping the large-scale structure (LSS) of the Universe with unparalleled statistical precision.
As these observations extend toward smaller scales and higher redshifts, the limiting factor in our ability to extract robust cosmological information is increasingly the accuracy of the underlying theoretical models for non-linear structure formation.

The theoretical modelling challenge has driven significant advances in computational cosmology, particularly through large-scale N-body simulations that can accurately capture the non-linear gravitational evolution of dark matter. Modern simulation suites, such as the MillenniumTNG simulations \citep{MillenniumTNG}, the AbacusSummit simulations \citep{AbacusSummit}, and more recent efforts like the Euclid Flagship simulations \citep{Euclid_flaghsip}, provide detailed predictions for matter clustering across a wide range of scales and redshifts. These suites are essential for developing and calibrating theoretical models required to interpret observational data and extract cosmological parameters.

Modelling the spatial distribution of matter alone is insufficient for interpreting galaxy survey data. Observationally, we measure galaxy positions in redshift space rather than real space, as the Doppler shifts induced by peculiar velocities modify distance estimates along the line of sight. This effect, known as redshift-space distortions (RSD), has been a subject of study since the 1970s \citep[e.g.][]{Jackson1972, Sargent1977}. The linear-theory framework established by \citet{Kaiser_1987} turned RSD into a cornerstone of LSS analyses. Rather than being a mere systematic effect, RSDs contain valuable cosmological information about the growth rate of structure; because peculiar velocities are generated by gravitational instability, they remain intimately connected to the underlying matter distribution.

The correct modelling of the cosmic peculiar velocity field is therefore paramount for two reasons: first, to accurately account for the systematic effects of RSD on clustering measurements, and second, to extract the growth rate information encoded in the velocity-density correlations. 
However, accurate measurement of velocity statistics from N-body simulations presents significant technical challenges. Volume-weighted velocity fields, which provide the most direct connection to the continuous velocity field, are difficult to estimate reliably from discrete particle distributions \citep[see, e.g.,][]{BernardeauEtal1997, Scoccimarro_2004, Hahn_2015}. As demonstrated by \cite{ZhengZhangJing2015, ZhangZhengJing2015}, standard estimation methods suffer from intrinsic sampling artifacts that can bias velocity statistics, particularly on small scales where shot noise becomes significant. These systematic effects can propagate into theoretical predictions and compromise the accuracy of RSD models, making careful convergence studies with varying resolution essential for robust results.

One of the key quantities in RSD modelling is the velocity divergence field $\theta = \nabla \cdot \mathbf{v}$, specifically its auto-power spectrum $\Ptt$ and its cross-power spectrum with the density field $\Pdt$ \citep{Scoccimarro_2004}.
For this reason, several theoretical frameworks have been developed to model $\Ptt$ and $\Pdt$ in the non-linear regime. \cite{Jennings2011}, with updates in \cite{Jennings2012}, proposed a universal fitting function relating the density and velocity divergence power spectra, with redshift evolution captured through the linear growth factor. However, their analysis was limited to a single simulation resolution, preventing assessment of numerical convergence. \cite{Hahn_2015} used fully converged measurements from simulations with different resolutions, though their analysis was restricted to $z = 0$.

A more comprehensive approach was developed by \cite{Bel_2019}, who provided accurate fitting functions that showed excellent agreement with simulation data over their calibration range. These fitting functions, however, rely on $\sotto$---the rms variance of linear density fluctuations smoothed on 8 Mpc/$h$ scales---as their measure of clustering amplitude. As demonstrated by \cite{Sanchez20} and further supported by \cite{Forconi_2025}, the explicit $h$-dependence in the smoothing scale prevents $\sotto$ from properly capturing the physical clustering amplitude. This leads to systematic errors in the non-linear power suppression in cosmologies with values of $h$ differing from the calibration suite.

This limitation reflects a broader issue in cosmological modelling: the widespread use of comoving coordinates scaled by $h$ can introduce spurious dependencies that compromise model accuracy when extrapolating across parameter space. A more robust approach involves using physical coordinates and amplitude measures that properly capture the underlying physics. Following the arguments of \cite{Sanchez20}, we advocate for using $\sdodici$—the rms variance smoothed on $12\,{\rm Mpc}$---which provides a reliable measure of clustering amplitude across different cosmological parameters.

Building on these insights and the evolution mapping framework developed in previous work \citep{Sanchez2021, Esposito_2024, AletheiaMass}, this paper presents improved fitting functions for \Pttanddt{} that address the limitations of existing models. 
Our approach incorporates two key improvements: first, we perform extensive convergence tests using simulations spanning multiple resolution levels to ensure reliable measurements across all scales and redshifts of interest; second, we adopt $\sdodici$ as the primary scaling parameter, eliminating spurious $h$-dependencies and ensuring accuracy across a broad cosmological parameter space.
These methodological advances provide the foundation for more robust theoretical predictions of galaxy clustering in redshift space, essential for maximising the scientific return from current and future LSS surveys.

The paper is organised as follows. We present and discuss the existing fitting functions available in the literature and present our improved formulas in \cref{sec:state-of-art}. In \cref{sec:sims} we introduce the three simulation suites that we use in this work, while in \cref{sec:methods} we describe how we use them to measure the non-linear $\Ptt$ and $\Pdt$. \cref{sec:results} contains our main results: the calibrated fitting formulas and their comparison with previous work. Finally, we give our conclusions and future perspectives in \cref{sec:conclusions}. We refer interested readers to \cref{apx:conv_tests} for convergence tests of our measurements.

\section{Fitting prescriptions and functional forms} \label{sec:state-of-art}

In this section, we present the state-of-the-art fitting prescriptions developed by \citet{Bel_2019} (hereafter B19), which serve as the baseline for our current study. We detail their functional forms and fitted parameters, and establish the conventions used throughout this work. The accuracy of these formulas, particularly in comparison to our new prescriptions, will be discussed in \cref{sec:results}. 

Throughout this section and the rest of the paper, we refer to $\theta\equiv-\nabla\!\cdot\!\mathbf{v}/(aHf)$ as the rescaled velocity divergence such that, at the linear level, the continuity equation reads $\theta = \delta$. All length-related quantities presented in this work are expressed in units of Mpc. For historical reasons, cosmologists often quote distances in units of $\Mpch$, where $h$ is the dimensionless Hubble parameter defined as $h \equiv H_0 / (100\,\mathrm{km\,s^{-1}\,Mpc^{-1}})$. However, as noted by \cite{Sanchez20}, using these $h$-dependent units can obscure the dependence of the linear matter power spectrum, $\PL$, (and consequently of the non-linear power spectrum) on $h$. We refer the reader to \citet{Sanchez20} and \citet{Forconi_2025} for a detailed discussion on the pitfalls of using $\Mpch$ units.

\subsection{Baseline fitting prescriptions} \label{subsec:Bel_fit}

\begin{table}
    \centering
    \caption{Fitting parameters for the B19 model as functions of
    $\sottom$, separated by the power spectrum they refer to. 
    }
    \label{tab:B19_params}
    
    \begin{minipage}{0.48\linewidth}
        \centering
        \begin{tabular}{lc}
            \hline
            \multicolumn{2}{c}{\boldmath{$P_{\theta\theta}$}} \\
            \hline
            $a_1$ & $-0.817 + 3.198\,\sotto$ \\
            $a_2$ & $\phantom{-}0.877 - 4.191\,\sotto$ \\
            $a_3$ & $-1.199 + 4.629\,\sotto$ \\
            \hline
        \end{tabular}
    \end{minipage}
    \hfill
    \begin{minipage}{0.48\linewidth}
        \centering
        \begin{tabular}{lc}
            \hline
            \multicolumn{2}{c}{\boldmath{$P_{\delta\theta}$}} \\
            \hline
            $1/k_\delta$ & $-0.017 + 1.496\,\sotto^2$ \\
            $b$          & $\phantom{-}0.091 + 0.702\,\sotto^2$ \\
            \hline
        \end{tabular}
    \end{minipage}
\end{table}

With the help of the DEMNUni simulations \citep{Carbone_2016}, which trace the structure formation of massive neutrino cosmologies at high resolution, \Bel{} demonstrated that the non-linear suppression of the correlation in the velocity divergence field depends strongly on the clustering amplitude, which they characterised in terms of $\sotto$. Interestingly, they found that the relevant parameter for correctly capturing the effect on non-linear evolution is the variance of the total matter field (including neutrinos), $\sottom$, rather than that of the cold dark matter alone, $\sottoc$, as is the case for the density field \citep{Castorina_2015}. Hereafter, we drop the subscript $\mathrm{m}$ and refer to the variance of the total matter field simply as $\sotto$.

\Bel{} proposed two fitting functions for $P_{\theta\theta}$ and $P_{\delta\theta}$; here, we focus on their most accurate prescriptions, which follow the functional forms:
\begin{align}
    P_{\theta\theta}(k) &= P^{\mathrm{Lin}}_{\theta\theta}(k)\, e^{-k(a_1 + a_2 k + a_3 k^{2})} \, , \label{eq:Bel_fit_tt} \\
    P_{\delta\theta}(k) &= \left[ P^{\mathrm{HF}}_{\delta\delta}(k) P^{\mathrm{Lin}}_{\theta\theta}(k) \right]^{1/2}
e^{-k/k_{\delta} - b k^{6}} \, , \label{eq:Bel_fit_dt}
\end{align}
where $P^{\mathrm{Lin}}_{\theta\theta}$ is the linear-theory velocity divergence power spectrum, which, with our definition of $\theta$, is identical to the matter one, $\PL$. $P^{\mathrm{HF}}_{\delta\delta}$ is the semi-analytic prediction for the non-linear matter power spectrum computed using the halo-fit prescription \citep{Halofit,Halofit_2}.
Here, the free parameters $a_1, a_2, a_3, k_{\delta}, b$ are fitted separately at each redshift; their dependence on $\sottom(z)$ is then modelled using linear or quadratic formulas. The best-fitting values are reported in \cref{tab:B19_params}.

With this parametrisation, \Bel{} reported a 3\% accuracy on scales $k \lesssim 0.7 \hMpc$. 
However, it is important to note that the only convergence tests performed by \Bel{} involved varying the grid size. As pointed out by \cite{ZhengZhangJing2015}, sampling artifacts introduced by estimating the velocity field from a mass-weighted distribution are independent of the grid size and depend only on the tracer sparsity (i.e., on the simulation resolution). 
The method of \Bel{}, based on the spherical smoothing of a Monte Carlo sampled Delaunay tessellation of the field, is precise but remains subject to resolution effects. We argue in \cref{sec:methods} and in more detail in \cref{apx:conv_tests} that, for their simulation setup, measurements become unreliable at the per cent level at scales as large as $k\sim0.2 \text{-} 0.3 \hMpc$.

Calibrating the fit in terms of $\sotto$ introduces an additional conceptual complication. When the shape of the linear power spectrum is held fixed, variations in the Hubble parameter $h$ should affect the clustering of matter only through their impact on the growth of structure, and hence on the overall amplitude of density fluctuations at a given redshift. By construction, however, $\sotto$ depends on $h$ not only through the growth factor but also through the definition of the smoothing scale itself, which is fixed to $8\,\mathrm{Mpc}/h$. As a result, changing $h$ modifies the physical scale over which fluctuations are smoothed, so that variations in $\sotto$ reflect both genuine changes in clustering amplitude and artificial changes in the smoothing radius. For this reason, $\sotto$, and more generally any amplitude parameter defined on an $h$-dependent scale, does not provide a clean measure of the physical clustering strength. Fitting prescriptions such as those of \Bel{}, which parametrise non-linear evolution in terms of $\sotto$, therefore inherit a spurious dependence on $h$, leading to unphysical and systematically biased predictions when varying the Hubble parameter. 

We address these limitations in the next section by proposing a new parametrisation within the Evolution Mapping framework \citep{Sanchez2021, Esposito_2024}, substituting $\sotto$ with $\sdodici$ to eliminate artificial $h$-dependencies.

\begin{table}
    \centering
    \caption{Cosmological parameters of our reference \LCDM{} model. This corresponds to the cosmology of the AletheiaMass simulation and of Model 0 in the Aletheia simulations. Note that there is a typo in the value of $n_\mathrm{s}$ reported in the equivalent table (Table 1) of \citet{Esposito_2024}.}
    \begin{tabular}{c c}
        \hline
         Parameter & Value \\
        \hline
         $\ob$ & 0.02244 \\
         $\oc$ & 0.1206  \\
         $\oDE$ & 0.3059      \\
         $\oK$ & 0    \\
         $h$ & 0.67   \\
        \hline
    \end{tabular}
    \begin{tabular}{c c}
        \hline
         Parameter & Value \\
        \hline
         $w_{0}$ & -1 \\
         $w_a$ & 0 \\
         $n_\mathrm{s}$ & 0.96 \\
          $\sigma_{12}(z=0)$ & 0.825 \\
          $A_\mathrm{s}$ & $2.127 \times 10^{-9}$\\
        \hline
    \end{tabular}
    \label{tab:LCDM_params}
\end{table}

\begin{table*}
\centering
\caption{Key parameters of the three simulation suites used in this work. Each suite is fully described in its own publication; we refer to the references listed in the last row for further details.}
\begin{tabular}{l|ccc}
\hline
& \textit{Aletheia} & \textit{AletheiaMass} & \textit{AletheiaEmu} \\
\hline
Cosmologies      & 9 variants (fixed $\Theta_s$, varied $\Theta_e$) & single, \LCDM{}, \cref{tab:LCDM_params} & 150 in Latin hypercube of $\Theta_s, \sdodici(z)$ \\
Box size [Mpc]   & 1492.5 & [180, 350, 700, 1400, 2800] & 1500 \\
Particles        & $1500^{3}$ & $1500^{3}$ / $2048^{3}$ & $2048^{3}$ \\
$\sdodici$ values & 0.343--0.825 (5 values per simulation) & 0.2--1.0 (10 values per simulation) & 0.2--1.0 (150 values in total) \\
Main use         & Growth history effect calibration & Main calibration & Performance evaluation \\
Reference        & \cite{Esposito_2024} & \cite{AletheiaMass} & \cite{AletheiaEmu} \\
\hline
\end{tabular}
\label{tab:simulation_summary}
\end{table*}

\subsection{Evolution–mapping–based prescriptions}
\label{subsec:our_fit}

Evolution mapping \citep{Sanchez2021} leverages a fundamental degeneracy in cosmological parameters by dividing them into two distinct categories: \textit{shape parameters} ($\shapep$), which define the shape of the linear matter power spectrum $\PL$, and \textit{evolution parameters} ($\evolp$), which only affect its amplitude at any given redshift. \cite{Sanchez2021} first showed that, when shape parameters are held fixed, the entire impact of the evolution parameters and redshift can be absorbed into a single parameter describing the clustering amplitude, such as $\sdodici(z)$; models matched in $\sdodici$ therefore display nearly identical non-linear matter statistics.

\cite{Esposito_2024} demonstrated that this degeneracy extends to the cosmic velocity field. Cosmologies with the same shape parameters and the same $\sdodici(z)$ exhibit highly consistent \Pttanddt{} up to mildly non-linear scales.

Crucially, this mapping is most robust when spectra are expressed in physical Mpc.
As discussed in \cref{subsec:Bel_fit}, \Bel{} parametrise the damping of linear modes with a dependence on $\sotto$. As a consequence, two cosmologies with identical $\shapep$ and equal \emph{physical} clustering amplitude but different $h$ values yield divergent damping predictions under the B19 model, despite simulations demonstrating that their actual $P_{\theta\theta}$ are nearly indistinguishable. By adopting $\sdodici$ and evaluating scales in physical Mpc, we eliminate this spurious $h$-dependence and preserve the exactness of the evolution mapping relation.

In addition to calibrating the models in physical units, we introduce a modified functional form. Renormalised perturbation theory \citep[RPT,][]{RPT} suggests that a non-linear spectrum can be viewed as the sum of (i) a \emph{damped} linear term, multiplied by a propagator that suppresses power on small scales, and (ii) an \emph{additive} mode-coupling contribution. Motivated by this insight, our proposed \emph{ansatz} supplements the exponential damping term from \Bel{} with an additive Gaussian contribution in $\ln k$.

Furthermore, we replace the linear matter power spectrum, $\PL$, in our \emph{ansatz} with its \textit{de-wiggled} counterpart, $\Pdw$, which accounts for the damping of the Baryon Acoustic Oscillation (BAO) feature due to non-linear evolution. Following the predictions of perturbation theory, we compute this as 
\begin{equation} \label{eq:P_dw}
    \Pdw = \PL G(k,z) + \Pnw \left [ 1-G(k, z)\right],
\end{equation}
where $\Pnw$ is the linear power spectrum with the BAO signal removed using the technique described in \cite{Hamann_2010}, and $G(k,z)$ is a Gaussian smoothing kernel
\begin{equation}
    G(k, z) = \exp\left[-\frac{1}{2}k^2\sv^2\right].
\end{equation}
This kernel depends on the linear velocity dispersion $\sv$, defined as
\begin{equation}
    \sv^2 = \frac{1}{6 \pi^2} \int_0^\infty \PL \, \diff k.
\end{equation}
This formulation yields a more accurate prediction of the BAO wiggles without requiring additional free parameters to characterise the non-linear damping.

While our damping function follows the functional form of \Bel{}, the additive Gaussian term introduces new coefficients, thereby extending the flexibility of the model to capture mode-coupling effects. Our complete functional form for the auto-spectrum reads:
\begin{equation}
\begin{aligned} \label{eq:our_fit}
P_{\theta\theta}(k,z) &=A\,\exp \Bigl[-\tfrac{\bigl(\ln k-\parmu\bigr)^{2}}{\parsigma}\Bigr] \\[6pt]
\;&+\; \Pdw\,\exp \bigl[-k\,\bigl(a_1 + a_2\,k +a_3 \, k^2\bigr)\bigr].
\end{aligned}
\end{equation}
The centre $\parmu$ and width $\parsigma$ are treated as constants with respect to $\sdodici$, while the amplitude $A$ and the coefficients $a_1, \, a_2, \, a_3$ are calibrated as simple functions of $\sdodici$.

For the cross-spectrum, we model $\Pdt$ as a function of $\Pdd$ and $\Ptt$. While at the linear level the continuity equation ensures that $P_{\delta\theta}^L(k) = \sqrt{P_{\delta\delta}^L(k)P_{\theta\theta}^L(k)}$, non-linear evolution suppresses the cross-correlation between the velocity and density fields. We thus adopt the following ansatz:
\begin{equation} \label{eq:cross_fit}
    \Pdt = \sqrt{\Pdd \Ptt} \, \, e^{-(c_1k+c_2k^2)},
\end{equation}
where $c_1$ and $c_2$ also follow a simple relation with $\sdodici$. For $\Pdd$, we rely on predictions from the {\tt Aletheia} emulator\footnote{\url{https://gitlab.mpcdf.mpg.de/arielsan/aletheia}} \citep{AletheiaEmu}, which achieves sub-per cent accuracy over the range of scales and $\sdodici$ considered here.

The explicit $\sdodici$-dependence of all parameters, together with a quantitative assessment of the fit, will be presented in \cref{sec:results}.

\section{Numerical simulations} \label{sec:sims}

We use three complementary suites of gravity-only \(N\)-body simulations---Aletheia, AletheiaMass, and AletheiaEmu---to measure the power spectra used to calibrate and validate our fitting formulas. All runs were performed with \Gadget{} \citep{G4}, starting from second-order Lagrangian perturbation theory (2LPT) initial conditions. These were generated using the paired-fixed technique of \citet{paired-fixed} to suppress cosmic variance. A concise overview of these suites is provided in \cref{tab:simulation_summary}; below, we summarise the features most relevant to this work.

\begin{itemize}
    \item Aletheia \citep{Sanchez2021,Esposito_2024}.  \\
    This suite comprises nine simulation pairs that share identical shape parameters but differ significantly in their growth histories. We report the cosmological parameters for the reference model in \cref{tab:LCDM_params}, and refer the reader to \citet{Esposito_2024} for details on the other cosmologies. 
    Each run evolves \(1500^{3}\) particles in a \(1492.5\;\mathrm{Mpc}\) box. Snapshots are saved at specific values of the clustering amplitude, \(\sdodici\), rather than at fixed redshifts. Consequently, corresponding snapshots across the different cosmologies have, by construction, identical linear matter and velocity-divergence power spectra while differing slightly in their non-linear growth histories---an ideal setup for isolating the impact of the growth rate on velocity statistics.

    \item AletheiaMass \citep{AletheiaMass}.  \\
    These simulations adopt a single cosmology (see \cref{tab:LCDM_params}) but vary both the box size ($180, 350, 700, 1400, \text{and } 2800 \Mpc$) and the number of particles (\(1500^{3}\) or \(2048^{3}\)). Each run starts at \(z = 99\) and outputs snapshots at ten intervals over the range $0.2 < \sdodici < 1.0$ (corresponding to \(-0.4 \lesssim z \lesssim 4.2\)). The wide dynamic range in mass and spatial resolution makes this suite ideal for convergence studies. As described in \cref{sec:methods} and \cref{apx:conv_tests}, we combine these different volumes to construct resolution-converged measurements of \(\Ptt\) and \(\Pdt\) across a broad range of \(k\) and \(\sdodici\), which serve as the primary calibration set for our fitting functions.

    \item AletheiaEmu \citep{AletheiaEmu}.  \\
    This large ensemble (100 training and 50 test simulation pairs) samples a Latin hypercube in the parameter space \((\oc, \ob, n_s, \sdodici)\). Each run evolves \(2048^{3}\) particles in a \(1500\;\mathrm{Mpc}\) box from 2LPT initial conditions at \(z = 99\) and outputs a single snapshot at a target $\sdodici$ within $0.2 < \sdodici < 1.0$. We use a subset of these simulations to assess the performance of our fitting formulas on models that are entirely independent of the calibration set.
\end{itemize}

Together, these three suites enable us to disentangle growth-history effects, verify numerical convergence, and assess the robustness of our fits across a broad range of cosmological parameters.

\begin{figure*}
    \centering
    \includegraphics[width=\textwidth]{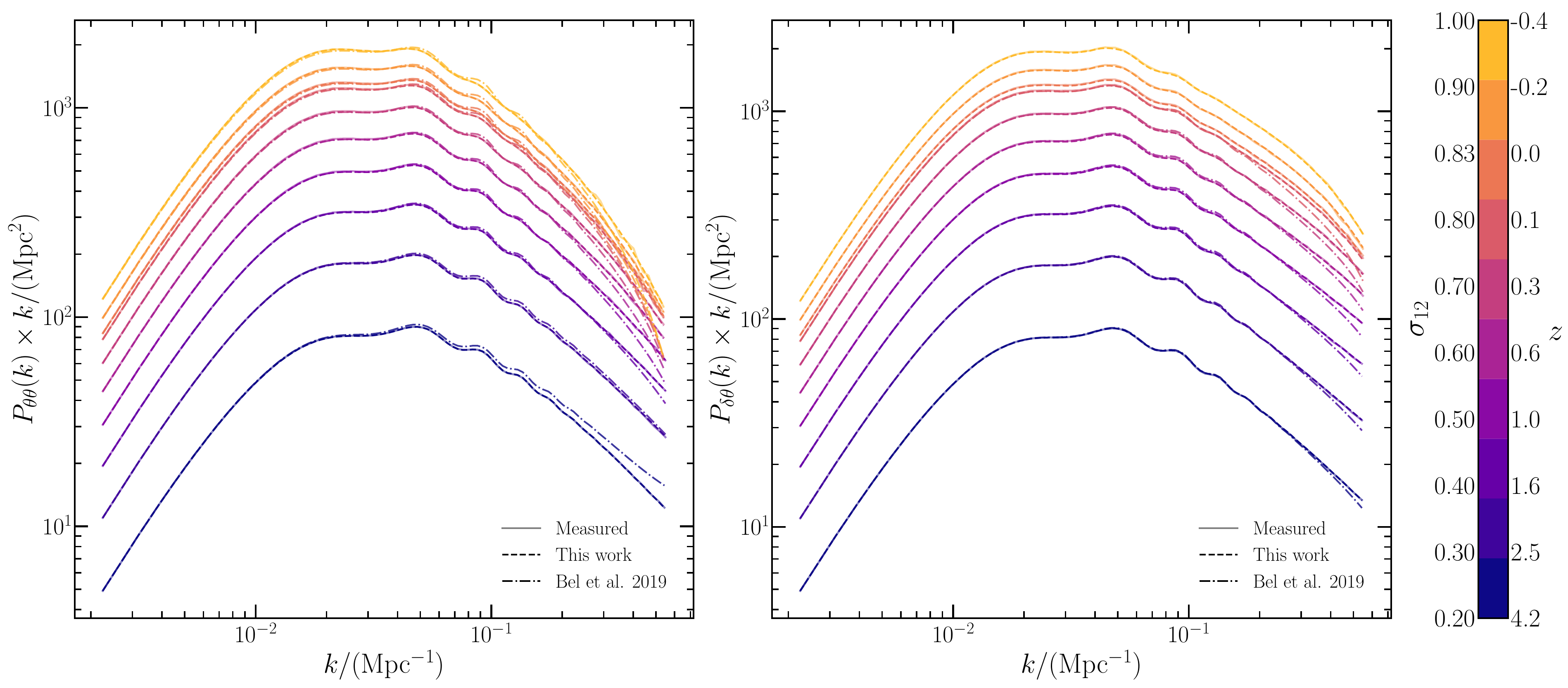}
    \caption{Velocity divergence auto- (left panel) and cross- (right panel) power spectra measured from the AletheiaMass simulations. The power spectra are multiplied by $k$ for a clearer display of the lines. Each colour indicates a different snapshot, with the corresponding values of $\sdodici(z)$ and $z$ indicated by the colour bar 
    . The measurements are shown with (shaded) solid lines, while the dashed lines indicate the fit obtained in this work and the dash-dotted lines display the prediction obtained with the fit of \Bel{}.}
    \label{fig:full_Pks}
\end{figure*}

\section{Estimating the density and velocity power spectra} \label{sec:methods}

One of the major challenges in measuring statistics of the cosmic velocity field from N-body simulations comes from the fact that massive particles trace the density field, not the velocity. While the lack of particles in an empty patch gives well-defined (low) densities, it provides no direct information about the velocity field in those regions. Even in areas where there are some particles, when integrating over the kernel used for the smoothing, most of the information comes from the location where particles reside, essentially resulting in mass-weighted averages, in contrast with the volume-weighted averages that are assumed in theoretical models.

To address this issue, a variety of methods have been developed over the years to interpolate the field in empty regions and uniformly sample the simulation volumes. A few examples include methods based on phase-space sheet interpolation \citep{Hahn_2015,Feldbrugge2024}, on the Delaunay tesselation of the simulation particles (\citealp{W&B1998,Schaap2000}; DTFE, \citealp{Cautun2011}) or on their Voronoi tesselation, be it by simply using the nearest particle method \citep[NP, ][]{Zheng2013,KodaEtal2014} or through smoothing a uniformly re-sampled field \citep[MC-Voronoi, ][]{Esposito_2024}. The latter is the method we choose for this work because of its good balance between accuracy and computational convenience. We provide here a brief overview of the method and refer the interested reader to the original paper for further details. 

Assuming that the velocity field is constant within each Voronoi cell, we re-sample the velocity field at the positions of a glass-like distribution of points, generated according to the recipe of \citet{Davila-Kurban2021}. 
As these points uniformly sample the volume, assigning them the velocity of the nearest $N$-body particle allows us to construct a volume-weighted field. We then smooth this field onto a $1024^3$ grid using a Cloud-in-Cell (CIC) assignment and deconvolve the window function in Fourier space.
We also evaluate a second mesh grid shifted by half a grid size to correct for aliasing through an interlacing technique \citep{Sefusatti_2016}. We use this smoothed velocity field to calculate its divergence in Fourier space and compute its auto-power spectrum $\Ptt$ and its cross-power spectrum $\Pdt$ with the density field, similarly smoothed on a 1024$^3$ CIC grid. The power spectra are binned logarithmically, up to the Nyquist frequency of the grid $k_\mathrm{Ny}=\pi N_\mathrm{grid}/L$, where $N_\mathrm{grid}$ is the size of the grid and $L$ the box size.

However, regardless of the interpolation method, the sparsity of tracers inevitably introduces a ``sampling artifact'' \citep{ZhangZhengJing2015, ZhengZhangJing2015}, limiting the reliability of our measurements to much smaller wavenumbers. When information at particle positions is used to infer velocities elsewhere, the effective smoothing scale depends on the local particle density and clustering rather than the grid size. To mitigate this and obtain reliable power spectra over a wide dynamic range, we implement a "patching" scheme using the AletheiaMass suite.
This provides simulations at two different particle resolutions for each box size, allowing us to determine a convergence scale, $\kcrop$, for each box by comparing the high- and low-resolution measurements. We define $\kcrop$ as the maximum wavenumber where the low- and high-resolution runs agree within $1\%$ (see \cref{apx:conv_tests} for the detailed convergence analysis). The final power spectra are constructed by stitching together measurements from different boxes: at each wavenumber $k$, we select the measurement from the largest available simulation volume that satisfies the convergence criterion $k < \kcrop$. Overlapping regions are combined using Gaussian smoothing to ensure spectral continuity. This conservative approach yields measurements fully converged up to $k \sim 0.56 \iMpc$, as shown \cref{fig:full_Pks}.

For the Aletheia and AletheiaEmu suites, which lack multi-resolution pairs, we determine the valid scales by calibrating against the AletheiaMass results. Since the sampling artifact is primarily driven by tracer density, we estimate $\kcrop$ for these simulations based on their mass resolution. This procedure is detailed in \cref{apx:conv_tests}; here we report the estimated minimum $\kcrop$ across all snapshots. We thus limit our analysis of the AletheiaEmu measurements to $k \lesssim 0.2 \, \iMpc$ and the Aletheia measurements\footnote{We note that ratios of power spectra are generally more robust to sampling artifacts than absolute measurements, as systematic errors largely cancel out \citep[e.g.][]{Esposito_2024}.} to $k \lesssim 0.18 \, \iMpc$. 

Finally, we perform a similar estimate for the DEMNUni suite used in \Bel{} and find that measurements of $\Ptt$ and $\Pdt$ at that resolution may be similarly affected by sampling artifacts at $k \gtrsim 0.15 \, \iMpc$. \citet{Stadler2025} showed that the resolution dependence of the method used in \Bel{} closely resembles that of a CIC kernel with size set by the mean inter-particle separation. Although this behaviour is not identical to that of the MC-Voronoi method, our preliminary comparisons indicate that the resulting $\kcrop$ values are similar, with MC-Voronoi performing slightly better.

\section{Results} \label{sec:results}

\subsection{Calibrating the fitting functions} \label{sec:calib}

In this section, we present the calibration of the functional form introduced in \cref{subsec:our_fit} and compare its performance against the baseline model of \Bel{} (\cref{subsec:Bel_fit}).

The calibration is performed in a Bayesian framework with \texttt{nautilus}\footnote{https://github.com/johannesulf/nautilus} \citep{nautilus}, which uses importance sampling to provide fast and accurate posterior estimates. We fit each snapshot of the AletheiaMass simulations (patched with the method described in \cref{sec:methods}) against the prescriptions provided in \cref{eq:our_fit,eq:cross_fit}. 

We perform our fits assuming a standard Gaussian likelihood and adopt a simplified, purely diagonal approximation for the covariance matrix. Specifically, we consider only the Gaussian sample variance, neglecting off-diagonal terms and non-Gaussian contributions arising from non-linear mode coupling. We adopt this approximation as an effective weighting scheme. This allows us to efficiently explore a parameter space characterised by strong degeneracies, a scenario where a sampling-based approach is significantly more robust than simple least-squares minimisation. While non-Gaussian contributions to the covariance are present on the scales considered here, their primary effect would be to inflate the formal parameter errors rather than substantially shift the maximum-likelihood values. Because our objective is to identify the best-fit parameters for our prescriptions rather than to estimate their posterior uncertainties rigorously, this simplified covariance approximation is sufficient for our purposes.

Under these assumptions, we approximate the error $\sigma_i$ in a bin $k_i$ containing $N_{\mathrm{m},i}$ modes as
\begin{equation}
    \sigma_i = \frac{P_{xy}(k_i)}{\sqrt{N_{\mathrm{m},i}/2}}\;,
\end{equation}
while the likelihood we adopt takes the form
\begin{equation}
\ln \mathcal{L}(\boldsymbol{\Theta})
= -\frac{1}{2} \sum_{i}
\left[ \frac{\left( P_{xy, i} - P^\mathrm{fit}_{xy, i}(\boldsymbol{\Theta}) \right)^2}{\sigma_i^2} \right]\;.
\end{equation}
Here, $P_{xy, i}$ represents the measured $\Ptt$ or $\Pdt$ in the bin $k_i$, and $P^\mathrm{fit}_{xy, i}(\boldsymbol{\Theta})$ represents the corresponding prediction from the fitting functions for a given set of parameters $\boldsymbol{\Theta}$. The priors on the parameters are flat and chosen to be non-informative. 

Note that the error we adopt corresponds to the standard sample variance. In our fixed-and-paired simulations, this variance is expected to be significantly reduced; however, the magnitude of this reduction has not been explicitly quantified for velocity-field statistics. Nevertheless, it is reasonable to expect a suppression similar to that observed for the matter power spectrum \citep[see][for a detailed analysis of fixing and pairing effects]{Villaescusa_Navarro_2018}. Since this suppression is strongest on large, linear scales, it primarily affects modes that are less critical for the non-linear fits performed in this work. For these reasons, we simply adopt this Gaussian error estimate throughout our analysis.

We calibrate the fitting functions up to $\kmax = 0.56 \iMpc$ to ensure the numerical convergence of our measurements. Even though the measurements of $\Pdt$ are converged up to a higher $k$, we limit the fit to the same scale for consistency. 

\begin{figure}
    \centering
    \includegraphics[width=\linewidth]{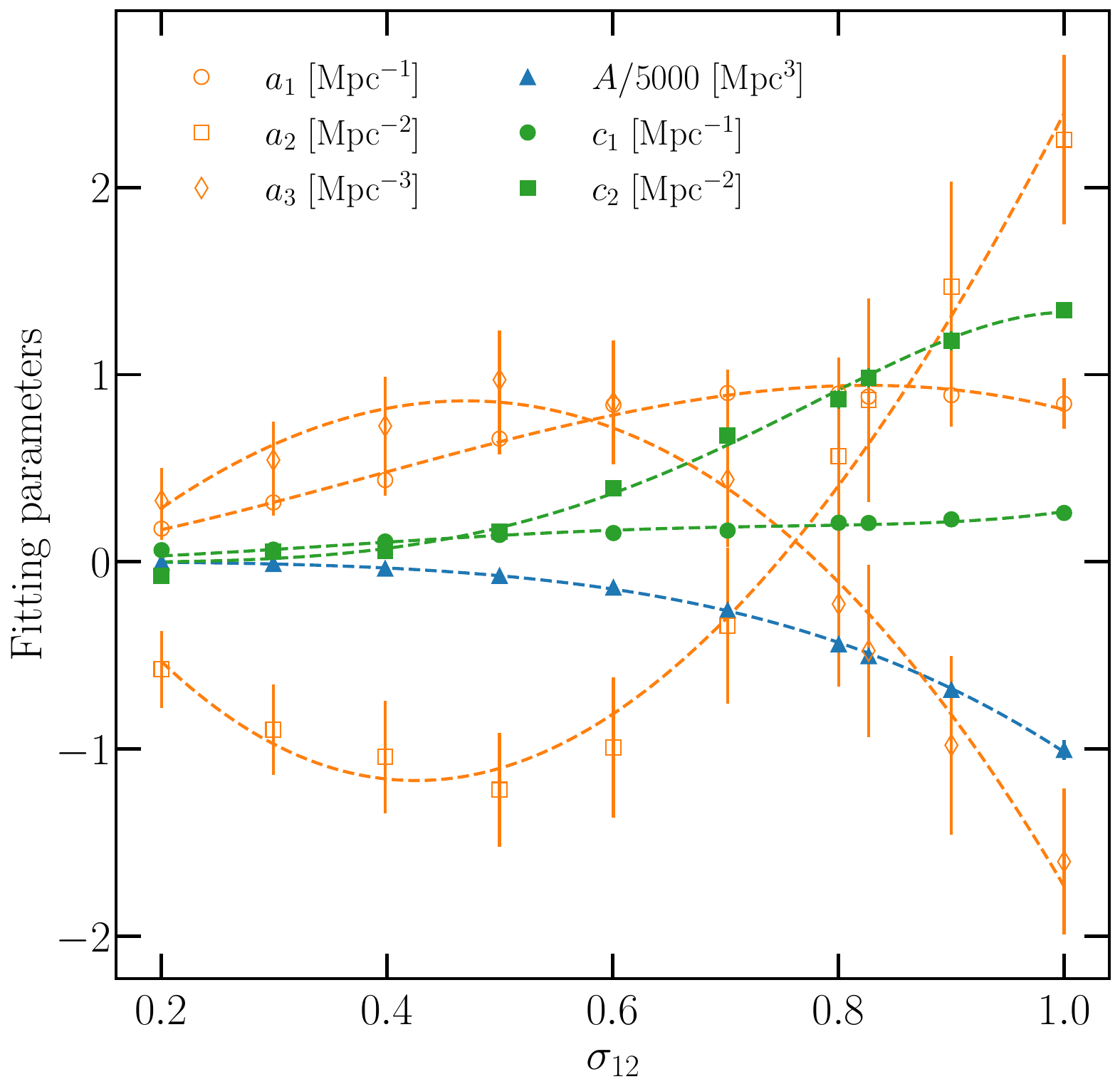}
    \caption{The dependence of the parameters for our \Pttanddt{} fitting prescriptions as a function of $\sdodici$. The dashed lines indicate the fitted relationships of \cref{eq:cosmo_dep} and \cref{eq:cosmo_dep_cross}. The three colours categorise the parameters: in blue and orange, the parameters of the additive and multiplicative terms in \cref{eq:our_fit}, respectively; in green, the parameters of \cref{eq:cross_fit}.}
    \label{fig:fit_params_vs_sigma_12}
\end{figure}

Let us first focus on $\Ptt$. We perform the fit in two steps: first, we marginalise over all other parameters to determine the best-fit values of $\parsigma$ and $\parmu$; then, we repeat the fit, this time fixing $\parsigma$ and $\parmu$ to these values for all snapshots. We adopt this approach because, in the first iteration, we find that these two parameters are only weakly sensitive to changes in $\sdodici$. This two-step process helps us obtain a more robust fit by reducing degeneracies. The remaining parameters show a stronger dependence on the degree of non-linear evolution, which we describe as a function of $\sdodici$. \Cref{fig:fit_params_vs_sigma_12} illustrates how these parameters depend on $\sdodici$ (blue and orange markers). We model their dependencies with a simple least-squares polynomial fit, yielding the following scaling relations:
\begin{equation} \label{eq:cosmo_dep}
\begin{aligned}
    a_1(\sdodici) &= \phantom{-}0.2895\,\sdodici \;+3.448\,\sdodici^{2} \;-2.925\,\sdodici^{3}\;, \\
    a_2(\sdodici) &= \phantom{-}1.176 \;- 11.4\,\sdodici \;+ 14.76\,\sdodici^{2} \;- 2.134\,\sdodici^{3}\;, \\
    a_3(\sdodici) &= -0.8032 \; +6.697\,\sdodici \;- 5.915\,\sdodici^{2} \;- 1.712\,\sdodici^{3}\;, \\
    A(\sdodici) &= \phantom{-}11.39 \;- 338.4\,\sdodici^2 \;- 4751\,\sdodici^4\;, \\
    \parsigma^2 &= \phantom{-}1.245\;, \\
    \parmu &= -3.156\;. \\
\end{aligned}
\end{equation}

\begin{figure*}
    \centering
    \includegraphics[width=\linewidth]{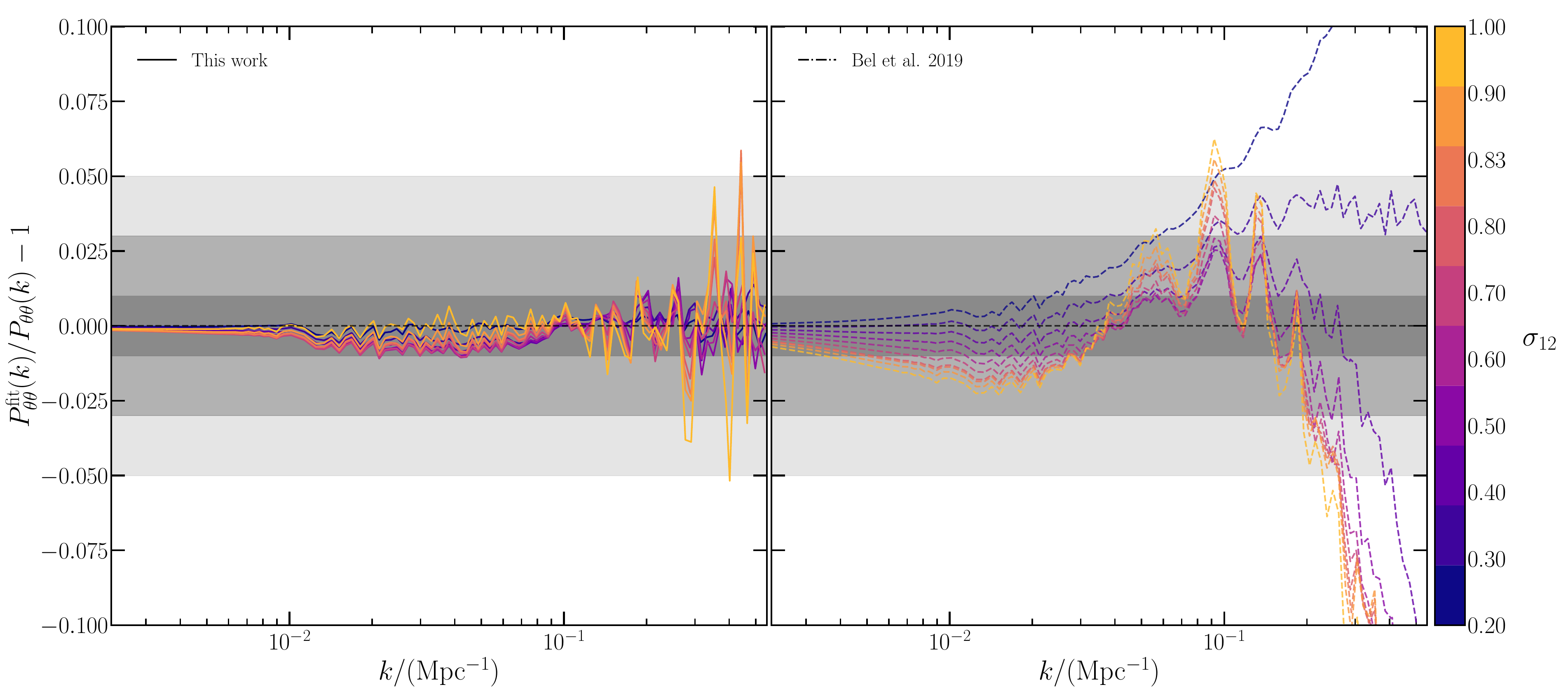}
    \caption{Accuracy of the fitting functions of \cref{eq:our_fit} (left panel) with parameters calibrated as in \cref{eq:cosmo_dep} (solid lines) compared to the $\Ptt$ of the calibration set (AletheiaMass). The residuals are shown as the relative differences of the predicted power spectra, $\Ptt^\mathrm{fit}$, with respect to the measured ones, $\Ptt$. The grey bands represent $1\%, \, 3\%$, and $5\%$ limits. The right panel shows the same comparison using the fitting functions of \Bel{} (dashed lines).}
    \label{fig:fit_vs_AleMass}
\end{figure*}

In the left panel of \cref{fig:fit_vs_AleMass}, we show the accuracy with which the functional form of \cref{eq:our_fit}, with parameters calibrated according to \cref{eq:cosmo_dep}, reproduces the $\Ptt$ measured from the AletheiaMass simulations. Overall, our model matches $\Ptt$ at the per cent level over a broad range of scales and clustering amplitudes.
For $k \lesssim 0.2 \iMpc$, the agreement is exceptionally tight across all values of $\sdodici$. At smaller scales ($k > 0.2 \iMpc$), the residuals for the earlier snapshots (low $\sdodici$) remain well-behaved and largely confined within the $1\%$ level. However, for late times ($\sdodici \gtrsim 0.8$), the measurements at small scales exhibit high-frequency oscillations that exceed $3\%$. These fluctuations reflect the larger sample variance of the smallest simulation box used for these physical scales. While this statistical noise dominates the measurements in this regime, the model predictions remain centred around the data with no obvious systematic bias, and are well within the data uncertainties (see \cref{apx:conv_tests} for details).

\begin{figure*}
    \centering
    \includegraphics[width=\linewidth]{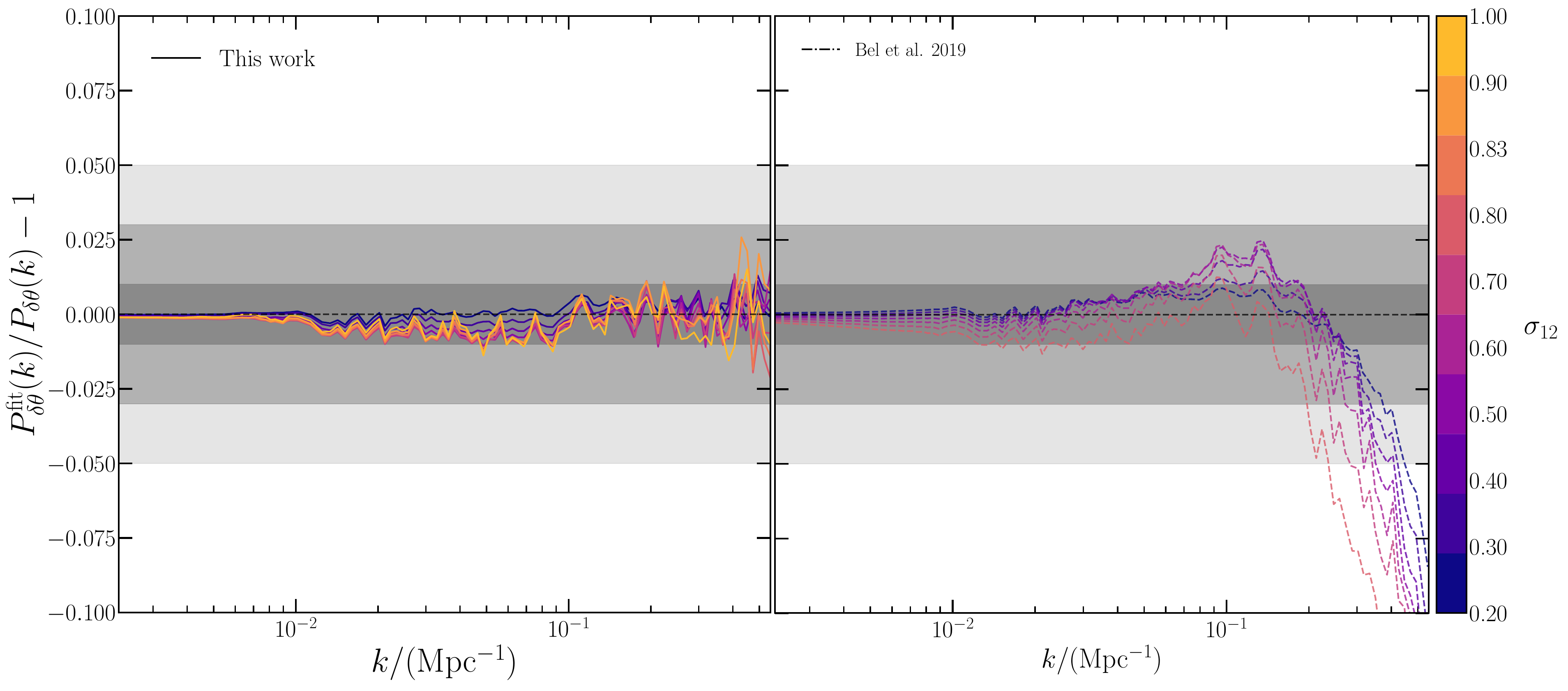}
    \caption{Accuracy of the fitting functions of \cref{eq:cross_fit} (left panel) with parameters calibrated as in \cref{eq:cosmo_dep_cross} (solid lines) compared to the $\Pdt$ of the calibration set (AletheiaMass). The residuals are shown as the relative differences of the predicted power spectra, $\Pdt^\mathrm{fit}$, with respect to the measured ones, $\Pdt$. The grey bands represent $1\%, \, 3\%$, and $5\%$ limits. The right panel shows the same comparison using the fitting functions of \Bel{} (dashed lines).}
    \label{fig:fit_vs_AleMass_cross}
\end{figure*}

For the $\Pdt$ calibration, we require estimates of both $\Ptt$ and $\Pdd$. For $\Ptt$, we use our fitted model with parameters given by \cref{eq:cosmo_dep}. As specified in Sec.~\ref{subsec:our_fit}, we describe $\Pdd$ using the predictions from the {\tt Aletheia} emulator \citep{AletheiaEmu}. We then fit our measurements against the functional form of \cref{eq:cross_fit}. As before, we model the parameter dependencies on $\sdodici$, finding the scaling relations
\begin{equation} \label{eq:cosmo_dep_cross}
\begin{aligned}
    c_1(\sdodici) &= \phantom{-}0.851\,\sdodici \;-1.328\,\sdodici^{2} \;+0.7462\,\sdodici^{3}\;, \\
    c_2(\sdodici) &= -0.1596\,\sdodici \;+ 4.262\,\sdodici^{2} \;- 2.769\,\sdodici^{3}\;. \\
\end{aligned}
\end{equation}
The left panel of \cref{fig:fit_vs_AleMass_cross} illustrates the quality of the $\Pdt$ fit when compared against the AletheiaMass measurements. In this case as well, we achieve per cent-level accuracy across all considered scales and $\sdodici$ values. Note that the $\Pdt$ measurements exhibit a smaller variance than those of $\Ptt$ due to their better convergence properties, which allow us to use the larger simulation boxes (with correspondingly smaller sample variance) down to smaller scales.

The right panels of \cref{fig:fit_vs_AleMass} and \cref{fig:fit_vs_AleMass_cross} show the accuracy of the \Bel{} fitting functions (dashed lines) compared to the $\Ptt$ and $\Pdt$ measured from the AletheiaMass simulations, respectively. A comparison with the left panels highlights the significant improvement in accuracy achieved by our new prescriptions. This improved performance can be attributed to several key differences in our approach:
\begin{enumerate}[wide=0pt]
\item While the formulas of \Bel{} were calibrated only up to $z \sim 1.5$, our prescriptions accurately predict $\Ptt$ and $\Pdt$ up to $z \sim 4$ ($\sdodici \sim 0.2$). Furthermore, cosmologies corresponding to negative redshifts are undefined in the \Bel{} model due to its reliance on \textsc{halofit} predictions (hence their absence in the right panel of \cref{fig:fit_vs_AleMass_cross}).
\item Owing to resolution-dependent sampling artefacts, the DEMNUni simulations used to calibrate the \Bel{} model render it reliable only up to $k \sim 0.1 \text{--} 0.2 \iMpc$, beyond which it systematically predicts less power. In contrast, our calibration relies on fully converged measurements across all scales considered in this work.
\item The inclusion of the additive term in \cref{eq:our_fit} allows our $\Ptt$ fitting function to better capture the overall shape of the power spectrum, leading to an improved description at large and intermediate scales.
\item Because our fits are calibrated on a de-wiggled linear power spectrum, the resulting residuals exhibit minimal BAO features. In contrast, the \Bel{} model, which takes $\PL$ as an input, does not adequately capture the non-linear damping of the BAO signal.
\item Finally, to avoid using quantities that depend explicitly on $h$, we parametrise the dependence on non-linear evolution in terms of $\sdodici$ rather than $\sotto$, and calibrate our models using $\iMpc$ units rather than $\hMpc$. This ensures our model correctly predicts velocity statistics across cosmologies with different Hubble parameters, a point we return to in \cref{sec:fit_vs_Ale}.
\end{enumerate}

\begin{figure*}
    \centering
    \includegraphics[width=\linewidth]{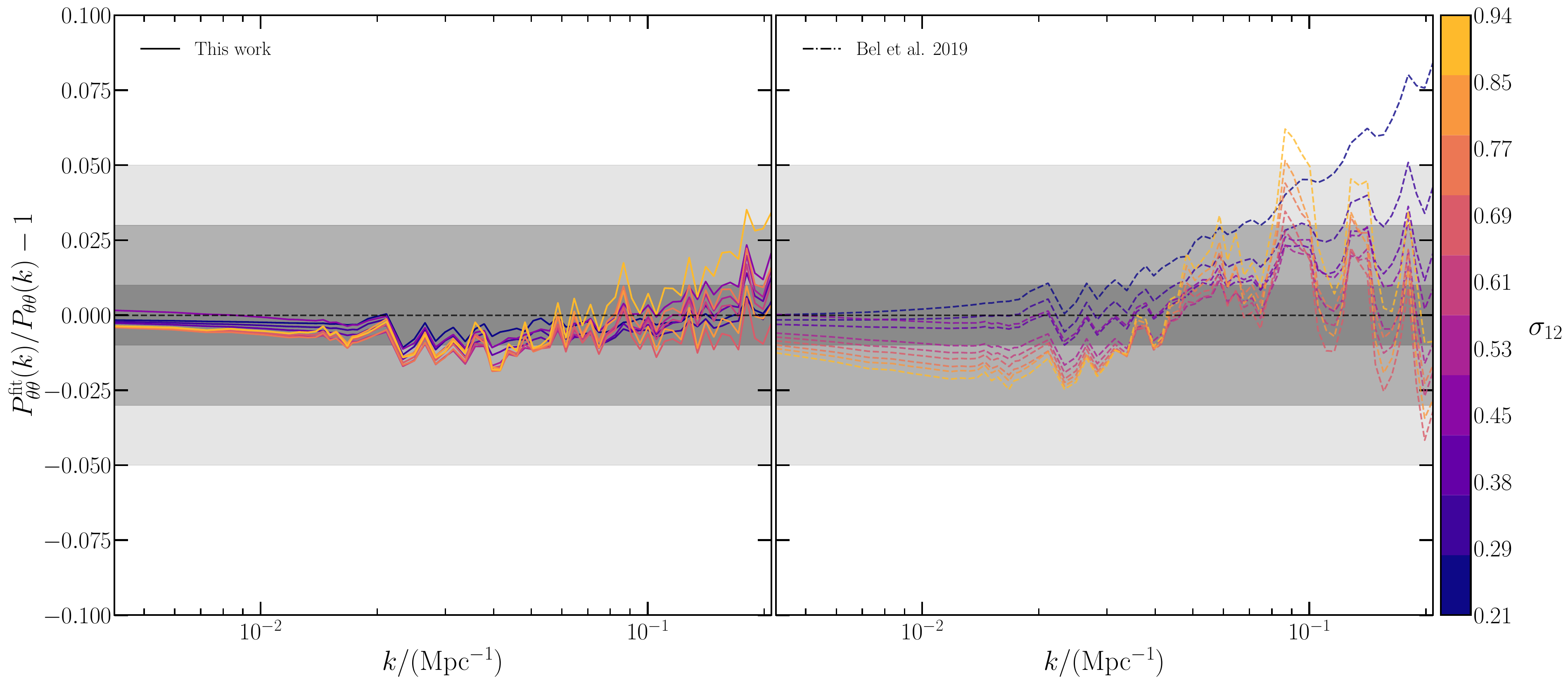}
    \caption{Accuracy of the fitting functions of \cref{eq:our_fit} (left panel) with parameters calibrated as in \cref{eq:cosmo_dep} (solid lines) compared to the $\Ptt$ of the AletheiaEmu simulations. The residuals are shown as the relative differences of the predicted power spectra, $\Ptt^\mathrm{fit}$, with respect to the measured ones, $\Ptt$. The grey bands represent $1\%, \, 3\%$, and $5\%$ limits. The right panel shows the same comparison using the fitting functions of \Bel{} (dashed lines).}
    \label{fig:fit_vs_AleEmu}
\end{figure*}

\begin{figure*}
    \centering
    \includegraphics[width=\linewidth]{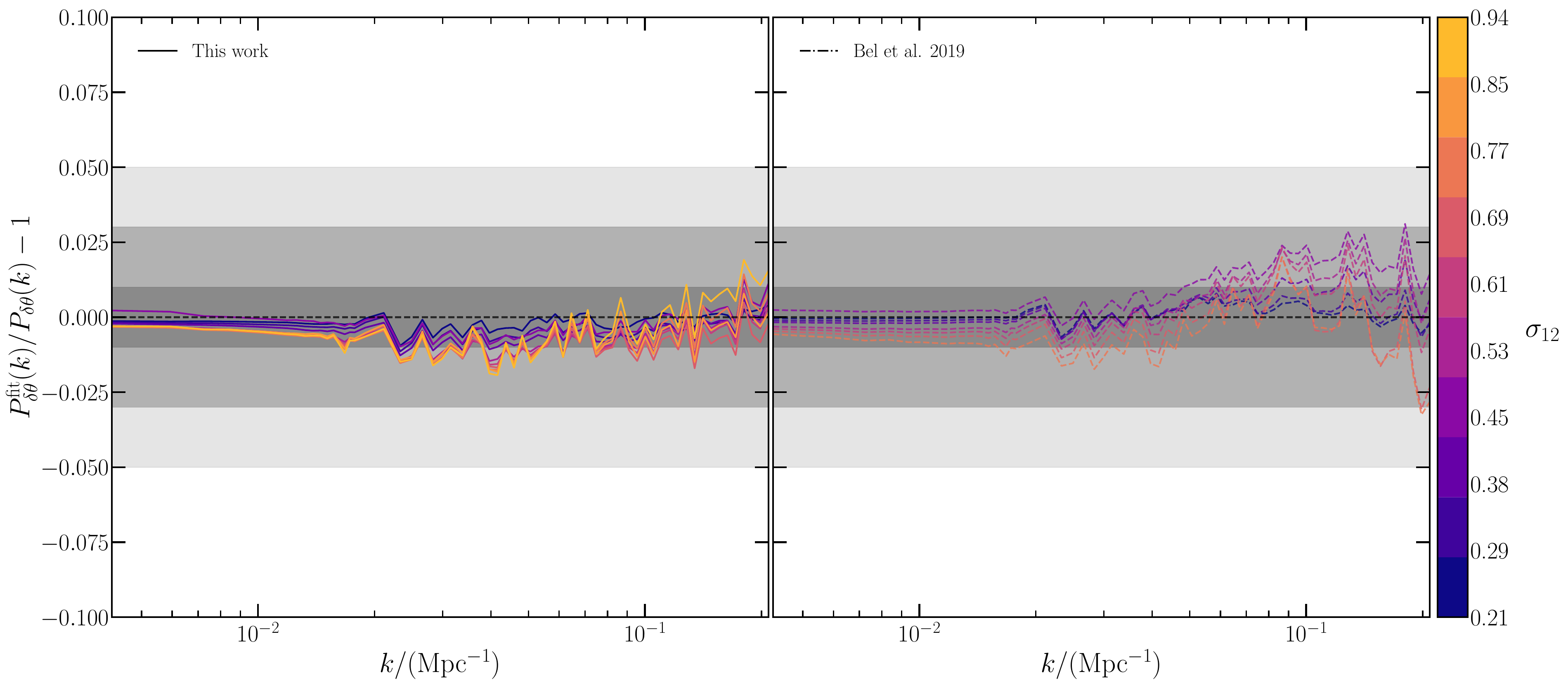}
    \caption{Accuracy of the fitting functions of \cref{eq:cross_fit} (left panel) with parameters calibrated as in \cref{eq:cosmo_dep_cross} (solid lines) compared to the $\Pdt$ of the AletheiaEmu simulations. The residuals are shown as the relative differences of the predicted power spectra, $\Pdt^\mathrm{fit}$, with respect to the measured ones, $\Pdt$. The grey bands represent $1\%, \, 3\%$, and $5\%$ limits. The right panel shows the same comparison using the fitting functions of \Bel{} (dashed lines).}
    \label{fig:fit_vs_AleEmu_cross}
\end{figure*}

We further assess the robustness of our models by comparing them against measurements of $\Ptt$ and $\Pdt$ from the AletheiaEmu simulations, which sample cosmologies with varying shape parameters. This comparison is shown in \cref{fig:fit_vs_AleEmu} and \cref{fig:fit_vs_AleEmu_cross}, where we restrict the analysis to $k < 0.21 \iMpc$ to suppress the impact of sampling artefacts (see \cref{apx:conv_tests}). Over this range, the agreement remains within $\sim 2\%$ at the smallest scales considered, with the exception of the snapshot at $\sdodici = 1$ in $\Ptt$, which shows deviations of up to $\sim 3\%$. Note that the predictions for $\Pdt$ (\cref{fig:fit_vs_AleEmu_cross}) generally exhibit better accuracy at small scales, as they are more resilient to sampling artefacts.

It is important to keep in mind that the estimates of the convergence scale $\kcrop$ presented in \cref{tab:crop_scales} are based on an interpolation. Consequently, we cannot conclusively determine whether the slight reduction in accuracy for the AletheiaEmu predictions reflects the fundamental limits of our fitting formulas across varying shape parameters, or if it simply arises from residual resolution effects in the test suite. Nevertheless, even in this regime, our prescriptions outperform those of \Bel{}, although the relative improvement is more modest on these scales.

\subsection{Impact of differing growth of structure histories}\label{sec:fit_vs_Ale}

\begin{figure*}
    \centering
    \includegraphics[width=\linewidth]{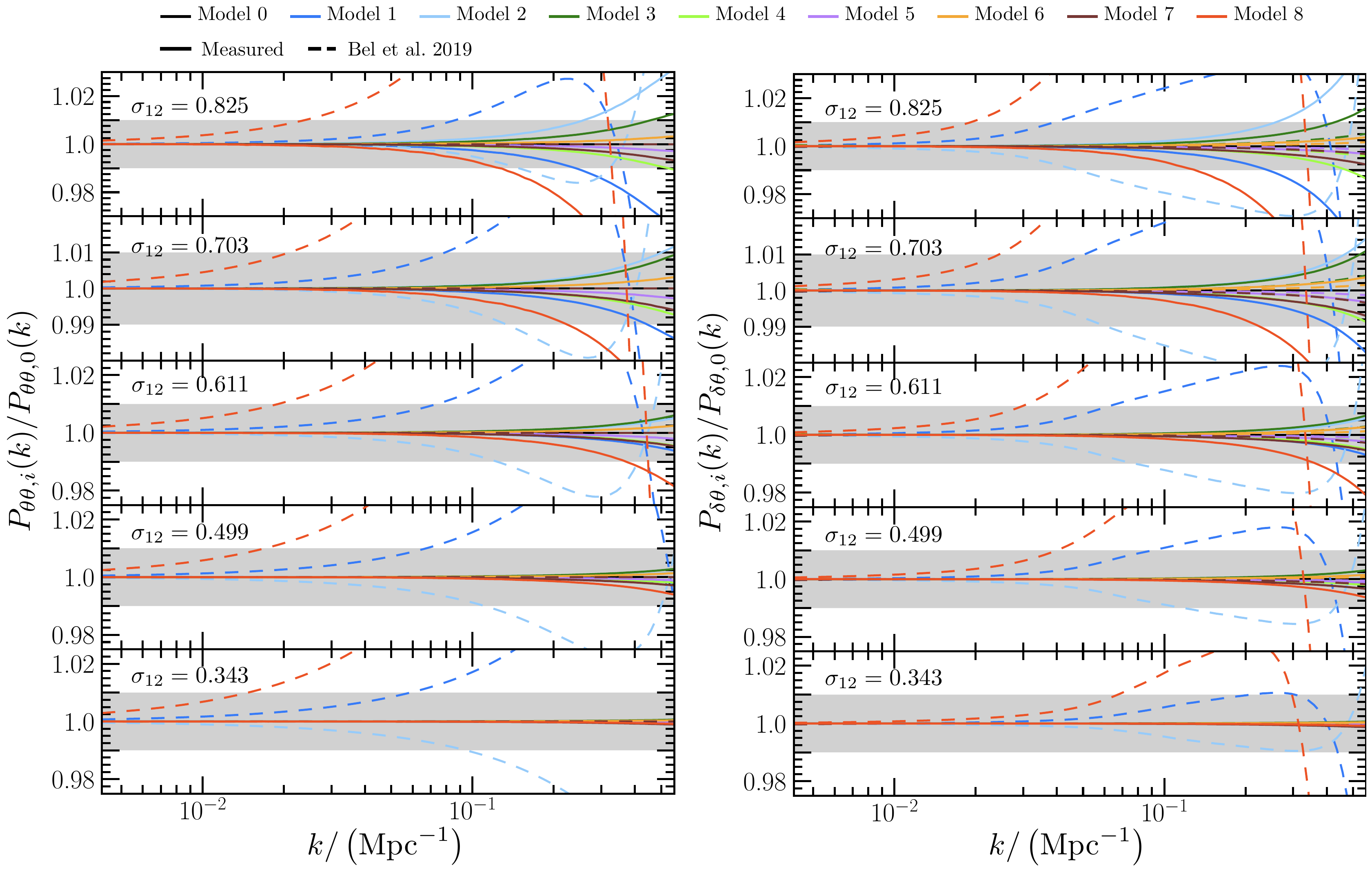}
    \caption{Ratios of velocity divergence auto-power spectra (left panel) and cross-power spectra with the density field (right panel) measured from the \textit{Aletheia} simulations relative to those of the reference cosmology (model 0). The different colours correspond to the various cosmologies detailed in Table 2 of \protect\cite{Esposito_2024}. The models are compared at snapshots corresponding to identical values of $\sdodici$. From bottom to top, the panels show five snapshots extracted at increasing values of $\sdodici$. The shaded grey regions indicate $1\%$ deviations from the reference model. The dashed lines show the corresponding ratios predicted by the \Bel{} prescriptions. While models parametrised by $\sdodici$ naturally preserve the evolution mapping relation, the dependence of the \Bel{} model on $\sotto$ introduces unphysical deviations for cosmologies with different values of $h$.}
    \label{fig:vPk_cPk_ratio_vs_Jul}
\end{figure*}

Our proposed prescriptions satisfy the evolution mapping relation by design. In practice, this means that cosmological models sharing the same shape parameters, $\shapep$, and evaluated at the same clustering amplitude, $\sdodici$, yield identical predictions for $\Ptt$ and $\Pdt$, regardless of their evolution parameters, $\evolp$. However, \cite{Esposito_2024} demonstrated that this perfect degeneracy breaks down slightly once strong non-linear structure formation sets in. In that regime, models reaching the same $\sdodici$ through different growth histories can exhibit small but measurable differences in their non-linear power spectra. 

{Recently, \citet{AletheiaEmu} and \citet{AletheiaMass} have successfully modelled these deviations for matter and halo statistics by introducing a parameter that captures the integrated recent structure growth history (specifically, a weighted average of $\Omega_{\mathrm{m}}/f^2$ evaluated over the recent evolution of $\sdodici$). However, as noted by \cite{Esposito_2024}, these deviations are minimal on the scales considered here, exceeding the $1\%$ level only for extreme cosmologies. Our goal is therefore to quantify the error incurred by neglecting these differences in the recent growth history when using our new fitting function for $\Ptt$ (with $\Pdt$ discussed below), which depends exclusively on the instantaneous value of $\sdodici$.

The Aletheia simulation suite provides an ideal testing ground for this purpose. It consists of simulations that share identical shape parameters, $\shapep$, and are analysed at snapshots corresponding to matching values of $\sdodici$ (hence at different redshifts), but span a wide range of evolution parameters, $\evolp$, and therefore possess markedly different structure growth histories.

We characterise the impact of differences in structure growth history by taking the ratio of the measured power spectra for each Aletheia model with respect to the reference \LCDM{} cosmology. These ratios are shown for \Pttanddt{} in the left and right panels of \cref{fig:vPk_cPk_ratio_vs_Jul}, respectively. Each panel shows, as solid lines, the ratios at the five values of $\sdodici$ available in the suite. 
Although the individual absolute measurements are affected by sampling artefacts for $k \gtrsim 0.18 \iMpc$ (see \cref{tab:crop_scales}), these systematic errors largely cancel out when considering ratios of power spectra. Consequently, we can safely extend this ratio analysis up to $\kmax = 0.56 \iMpc$, the maximum scale used to calibrate our fitting functions.

The left panel of \cref{fig:vPk_cPk_ratio_vs_Jul} shows that for most models, and over the full range of $\sdodici$ probed, the error introduced by neglecting differences in structure growth history is limited to the $\sim 1\%$ level. Only three models display significantly larger deviations: an Einstein--de Sitter model ($h=0.38$, model 8), and two models with exceptionally low ($h=0.55$, model 1) and high ($h=0.79$, model 2) values of the Hubble parameter. While these cases are useful for illustrating the breakdown of exact evolution mapping, they correspond to extreme growth histories that are ruled out by current observations. We therefore conclude that for most cosmologies of practical interest, growth history variations are subdominant in the total error budget of our $\Ptt$ fitting functions. We leave the explicit modelling of these residual deviations to future work---specifically, to the development of a dedicated velocity emulator within the \texttt{Aletheia} framework.

The right panel of \cref{fig:vPk_cPk_ratio_vs_Jul} shows the corresponding ratios for $\Pdt$. The overall magnitude of the deviations is comparable to that of $\Ptt$, and similar conclusions apply. There is, however, an important distinction: our fitting function for $\Pdt$ explicitly depends on the matter power spectrum, $\Pdd$, predicted by the \texttt{Aletheia} emulator. Because this emulator already incorporates the dependence of $\Pdd$ on structure growth history, part of this information is implicitly transferred into our $\Pdt$ description. This indirect sensitivity further mitigates the error introduced by omitting explicit growth-history parameters in our cross-spectrum model.

Finally, \cref{fig:vPk_cPk_ratio_vs_Jul} illustrates the problems associated with using $h$-dependent units or $\sotto$ statistics of the non-linear density and velocity fields. The dashed lines in both panels show the predicted power spectra ratios obtained using the prescriptions of \Bel{}. For any models whose value of $h$ matches the baseline adopted for the calibration in \Bel{}, the predicted $\Ptt$ is identical to the reference model (a ratio of exactly unity). However, models with different $h$ values exhibit a strong, systematic, and unphysical offset. As the models of \Bel{} are parametrised by $\sotto$, they retain a spurious $h$-dependence tied to the $8\,\mathrm{Mpc}/h$ smoothing scale, even when the models share identical linear physics. For $\Pdt$, these deviations are somewhat mitigated because their reliance on \textsc{halofit} for $\Pdd$ partially compensates for the effect, introducing a mild sensitivity to growth history similar to our approach.

While the cosmological models tested here are intentionally extreme, they clearly highlight the limitations of current standard practices. Even moderate variations in $h$ lead to non-negligible inaccuracies in prescriptions like those of \Bel{}. A simple re-parametrisation of the fitting functions of \Bel{} in terms of $\sdodici$ would immediately resolve this unphysical scaling, collapsing the dashed lines to unity. 
Overall, this demonstrates the significant systematic biases introduced by models that depend explicitly on $h$, strongly supporting our choice to use physical units ($\iMpc$) and $\sdodici$ for high-precision large-scale structure analyses.

\section{Conclusions} \label{sec:conclusions}

In this work, we have presented a new set of fitting functions for the velocity divergence auto- and cross-power spectra, $\Ptt$ and $\Pdt$, calibrated on high-resolution $N$-body simulations and designed to accurately describe the mildly non-linear regime relevant for RSD modelling. Our approach builds upon the prescriptions of \Bel{}, introducing several targeted modifications aimed at enhancing both robustness and the range of applicability.

Using the AletheiaMass simulations, we carefully quantify numerical convergence and mitigate sampling artefacts through a conservative patching strategy, allowing us to calibrate our fitting functions up to $k \simeq 0.56\,\mathrm{Mpc}^{-1}$. Within this range, we achieve per cent-level accuracy for both $\Ptt$ and $\Pdt$ over a wide span of $\sdodici$ values. When tested against the independent AletheiaEmu simulations, which probe a broad range of cosmological models with different shape parameters, the agreement remains at the $1$ -- $2\%$ level on scales where the measurements are robust, systematically outperforming existing fitting prescriptions.

We further quantify the impact of deviations from perfect evolution mapping induced by differences in structure growth history. We find that, for most cosmologies of practical interest, neglecting this effect introduces errors at or below the per cent level. Larger deviations arise only for extreme growth histories that are already strongly disfavoured by current observations. For $\Pdt$, part of this dependence is implicitly captured through the use of the {\tt Aletheia} emulator for $\Pdd$, which further reduces the impact of different growth histories. These results validate the use of $\sdodici$ as the sole parameter for our fitting functions, while clearly identifying the regime in which more complex modelling becomes necessary.

A key result of this paper is the demonstration that the choice of parametrisation and units plays a critical role in the robustness of fitting functions for velocity statistics. By explicitly testing models that share identical linear power spectra and identical values of $\sdodici$ but differ in their growth histories, we have shown that fitting functions expressed in $h$-dependent units introduce spurious, unphysical dependencies on the Hubble parameter. In particular, we find that fits relying on quantities such as $\sotto$ or scales expressed in $\mathrm{Mpc}/h$ can lead to systematic offsets when applied to cosmologies with different values of $h$, even when the underlying linear clustering and non-linear amplitude are identical. These systematic biases do not reflect underlying physical differences, but are driven entirely by the use of $h$-dependent units.

Our results provide strong support for adopting physical units and for parametrising non-linear evolution in terms of $\sdodici$ rather than $\sotto$. The former directly traces the clustering amplitude at the redshift of interest and preserves the physical degeneracies implied by evolution mapping, while avoiding spurious dependencies on the Hubble parameter. This choice is shown to be essential for achieving per cent-level accuracy across different cosmologies and for ensuring that deviations from perfect evolution mapping can be interpreted as the effect of true deviations in structure growth histories rather than systematic errors of the modelling.

The fitting functions presented here are directly relevant for the analysis of current and upcoming galaxy surveys, where accurate RSD modelling requires precise predictions for velocity statistics beyond linear theory. By providing per cent-level predictions for $\Ptt$ and $\Pdt$ on mildly non-linear scales, our results offer a practical and robust ingredient for cosmological parameters inference from galaxy surveys.

Looking ahead, the primary limitation to further progress lies in the accuracy with which velocity statistics can be measured from $N$-body simulations. A natural next step is therefore the construction of an emulator for $\Ptt$ and $\Pdt$, analogous to the existing {\tt Aletheia} emulator for the matter power spectrum \citep{AletheiaEmu}. Such an emulator would enable higher accuracy and increased robustness across a broad cosmological parameter space. As demonstrated for $\Pdd$, explicitly modelling this dependence captures residual deviations from exact evolution mapping, directly improving theoretical accuracy. A dedicated velocity emulator would thus provide a unified, high-precision model for both density and velocity fields, essential for next-generation galaxy clustering analyses.

\section*{Acknowledgements}

We would like to thank Carlos Correa, Sofia Contarini, Andrea Fiorilli, Jiamin Hou, Soumadeep Maiti, Alejandro P\'erez Fern\'andez, and Lukas Schwörer for their help and useful discussions. This research was supported by the Excellence Cluster ORIGINS, which is funded
by the Deutsche Forschungsgemeinschaft (DFG, German Research Foundation) under Germany’s Excellence Strategy - EXC-2094 -
390783311.

\section*{Data Availability}

The simulation data underlying this article will be shared on reasonable request to the corresponding authors.



\bibliographystyle{mnras}
\bibliography{bibliography} 



\appendix
\section{Convergence tests} \label{apx:conv_tests}

Estimating the volume-weighted velocity field from $N$-body simulations is challenging because the field is only sampled at particle positions. Evaluating the velocity at a location $x$ that does not coincide with a particle requires interpolating from the velocities of nearby tracers at positions $x_i$. This interpolation inevitably introduces a velocity sampling artifact, first investigated in \citet{ZhangZhengJing2015} and \citet{ZhengZhangJing2015}. A useful way to characterise this effect is through the deflection field, $\mathbf{d}(x) = x - x_{\rm nn}$, where $x_{\rm nn}$ is the position of the nearest particle. The reconstructed velocity field is effectively smoothed over scales of order $|\mathbf{d}(x)|$.

This velocity sampling artifact affects all velocity-field estimators to some extent, including advanced reconstruction techniques such as DTFE, natural-neighbour methods, and the MC--Voronoi estimator used in this work. Regardless of the interpolation method, the underlying limitation is the sparsity of the tracers, and the dominant parameter controlling the effect is the particle number density. We quantify this density using the resolution $R = N_\mathrm{p}/L^3$, where $L$ is the box size and $N_\mathrm{p}$ is the number of particles.

To assess the impact of this effect on our measurements, we perform a dedicated convergence test based on the AletheiaMass simulations. This suite provides runs with the same cosmology but two particle counts---$1500^3$ (low-resolution, LR) and $2048^3$ (high-resolution, HR)---across five different box sizes. These paired configurations allow a direct test of resolution effects over a broad dynamical range.

\begin{figure}\centering\includegraphics[width=\linewidth]{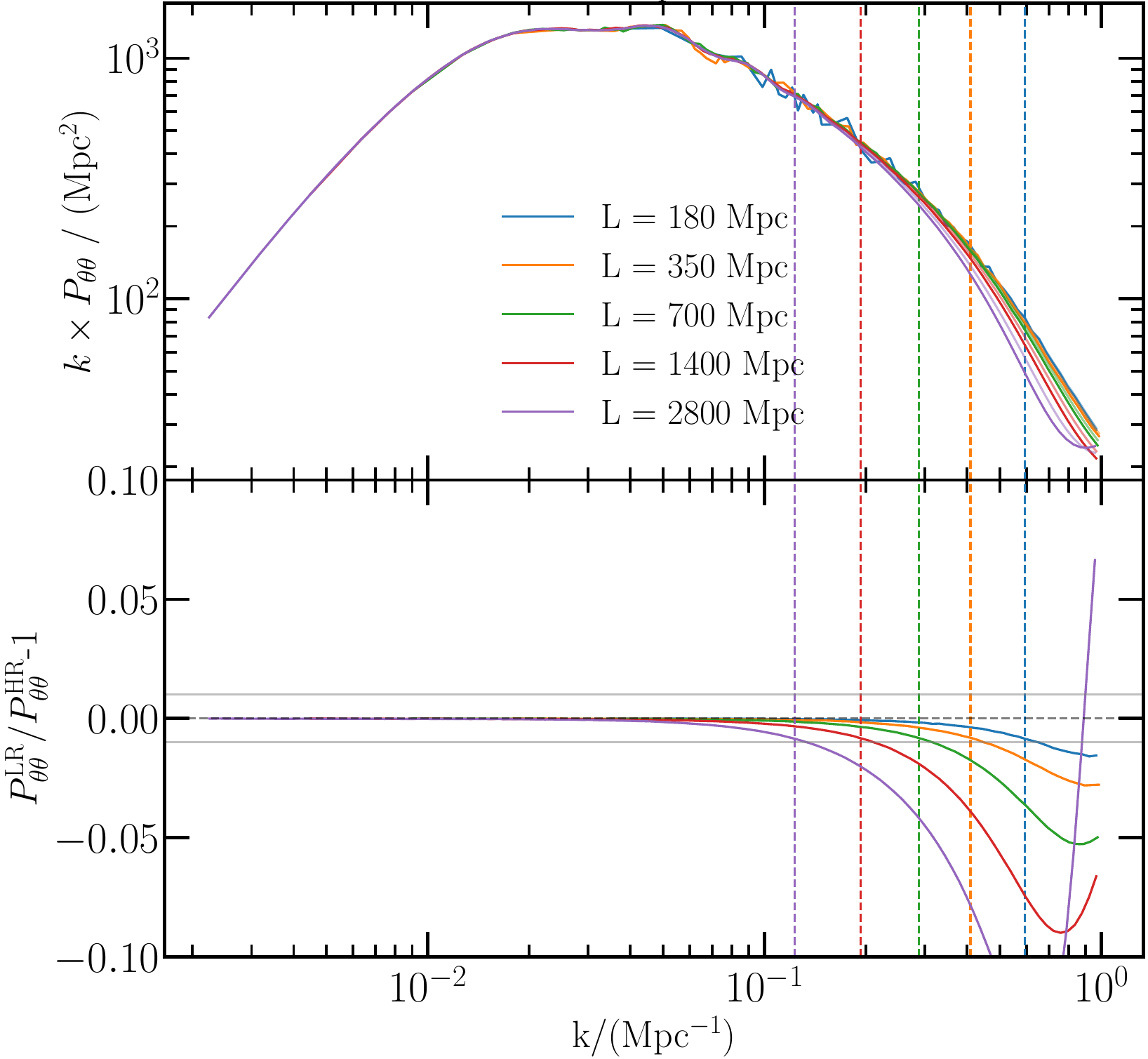}
    \caption{Illustration of the procedure used to determine the crop scales $\kcrop$ for each box size. As an example, we show $\Ptt$ at $z=0$. Top: Measurements of $k\,\Ptt$ from the HR simulations for all box sizes (solid lines). The corresponding LR measurements are shown as shaded lines and closely overlap with the HR curves. Vertical lines mark the inferred $\kcrop$ for each box, with colours matching those of the spectra. Bottom: Ratios of LR to HR measurements. The horizontal grey lines indicate a $1\%$ deviation from unity. The crop scale $\kcrop$ for each box is defined as the wavenumber at which the ratio first crosses the $1\%$ threshold; these scales are again indicated by the vertical coloured lines.}
    \label{fig:cropping_procedure}
\end{figure}

\begin{figure}\centering\includegraphics[width=\linewidth]{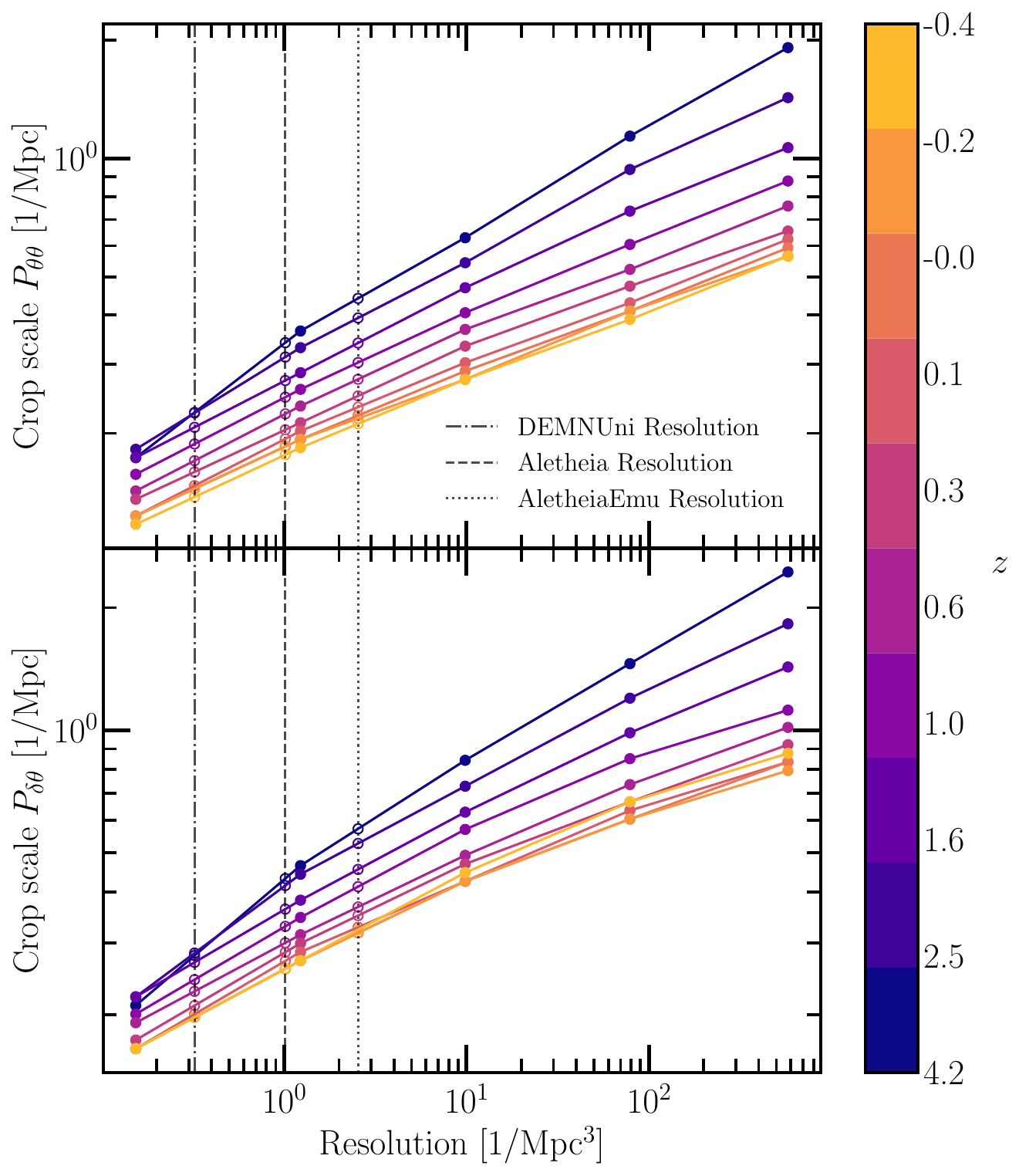}
    \caption{Crop scales $\kcrop$ for $\Ptt$ (top) and $\Pdt$ (bottom) measured from the AletheiaMass simulations as a function of resolution $R$. Different colours correspond to different redshifts. The vertical dash–dotted, dashed, and dotted lines mark the resolutions of the DEMNUni, Aletheia, and AletheiaEmu simulations, respectively. Empty circles denote the interpolated crop scales obtained by intersecting these vertical lines with the solid coloured curves, providing the estimated $\kcrop$ for each simulation at the redshifts available in AletheiaMass.}
    \label{fig:estimate_crop_other_sims}
\end{figure}

For each box and snapshot, we measure the ratio of the LR to HR power spectra and identify the wavenumber $\kcrop$ at which this ratio first deviates from unity by more than $1\%$. We take this $\kcrop$ as the maximum wavenumber at which the LR simulation is converged. An example of this procedure is shown in \cref{fig:cropping_procedure}. The LR--HR comparison probes the convergence of the LR runs. Applying these cuts to the HR measurements is therefore conservative: the HR runs remain converged to smaller scales, but this approach avoids model-dependent extrapolations. For this reason, whenever we quote $\kcrop$ as a function of resolution for the AletheiaMass suite, the associated resolution values correspond to the LR ($1500^3$) simulations. 

As this velocity sampling effect mainly depends on the particle density, the measured crop scales can be interpreted as a function of the resolution $R$. In \cref{fig:estimate_crop_other_sims} we show the dependence of $\kcrop$ on $R$, which increases monotonically with resolution: simulations with more particles experience smaller deflection distances and therefore retain accuracy down to larger wavenumbers. This scaling enables us to estimate $\kcrop$ for simulations lacking an equivalent convergence test. We infer the crop scales for the DEMNUni, Aletheia, and AletheiaEmu simulations by computing their respective resolutions and interpolating in log-space. The inferred values, together with the directly measured ones, are reported in \cref{tab:crop_scales}.

Finally, to minimise sample variance in small volumes while preserving accurate small-scale information, we construct a patched power spectrum for each snapshot using the computed crop scales of the AletheiaMass simulations. At every wavenumber $k$, we select the measurement from the largest simulation box that satisfies $k < \kcrop^L$, where $\kcrop^L$ is the crop scale associated with the LR simulation of box size $L$. This ensures that all retained data lie within a demonstrably converged regime while maximising the accessible range of scales.

\begin{table*}
    \centering
    \begin{tabular}{l|cccccccccc|c}
    \hline
    \hline
     & \multicolumn{10}{c|}{Crop scales \([\mathrm{Mpc}^{-1}]\)} & \multirow{3}{*}{$R$ [Mpc\(^{-3}\)]} \\
       $\sdodici$ & $0.2$ & $0.3$ & $0.4$ & $0.5$ & $0.6$ & $0.7$ & $0.8$ & $0.82$ & $0.9$ & $1.0$ & \\
      Redshift & $4.2$ & $2.5$ & $1.6$ & $1.0$ & $0.6$ & $0.3$ & $0.1$ & $0$ & $-0.2$ & $-0.4$ & \\
    \hline
    
    \multirow{5}{*}{AletheiaMass}
     & 1.9165 & 1.4297 & 1.0666 & 0.8773 & 0.7577 & 0.6544 & 0.6234 & 0.5930 & 0.5650 & 0.5650 & 578.70 \\
     & 1.1411 & 0.9386 & 0.7353 & 0.6049 & 0.5225 & 0.4738 & 0.4297 & 0.4093 & 0.4093 & 0.3897 & 78.72 \\
     & 0.6291 & 0.5434 & 0.4693 & 0.4053 & 0.3676 & 0.3333 & 0.3025 & 0.2880 & 0.2743 & 0.2743 & 9.84 \\
     & 0.3642 & 0.3303 & 0.2853 & 0.2588 & 0.2346 & 0.2128 & 0.2027 & 0.1930 & 0.1930 & 0.1838 & 1.23 \\
     & 0.1734 & 0.1821 & 0.1734 & 0.1573 & 0.1426 & 0.1358 & 0.1232 & 0.1232 & 0.1232 & 0.1173 & 0.15 \\
    \hline
    
    DEMNUni (est.)
     & 0.2260 & 0.2252 & 0.2071 & 0.1879 & 0.1704 & 0.1594 & 0.1472 & 0.1446 & 0.1446 & 0.1377 & 0.32 \\ 
    \hline
    
    Aletheia (est.)
     & 0.3401 & 0.3127 & 0.2725 & 0.2471 & 0.2241 & 0.2042 & 0.1936 & 0.1852 & 0.1852 & 0.1764 & 1.015 \\ 
    \hline
    
    AletheiaEmu (est.)
     & 0.4409 & 0.3931 & 0.3395 & 0.3027 & 0.2745 & 0.2490 & 0.2331 & 0.2220 & 0.2182 & 0.2114 & 2.54 \\ 
    \hline
    \hline
    \end{tabular}
    \caption{Crop scales for the $\Ptt$ measurements from the $N$-body simulations. The first five rows list the values obtained with the procedure described in \cref{apx:conv_tests} and correspond to the AletheiaMass simulations with $1500^3$ particles. The final column reports the resolution of each simulation, $R = N_\mathrm{p}/L^3$. The last three rows provide the crop scales estimated for the DEMNUni, Aletheia, and AletheiaEmu simulations, derived by interpolating the AletheiaMass results as illustrated in \cref{fig:estimate_crop_other_sims}. 
    }
    \label{tab:crop_scales}
\end{table*}


\bsp	
\label{lastpage}
\end{document}